\documentclass{article}
\usepackage{spconf,amsmath,graphicx}
\usepackage{color}
\usepackage[table,xcdraw]{xcolor}
\usepackage{amstext}
\usepackage{amsfonts}
\usepackage{float}
\usepackage[normalem]{ulem}
\useunder{\uline}{\ul}{}

\usepackage{enumitem}
\setlist{nosep, leftmargin=14pt}

\usepackage{mwe} 

\title{X-Ray2EM: Uncertainty-Aware Cross-Modality Image Reconstruction from X-Ray to Electron Microscopy in Connectomics}
\name{
\begin{tabular}{c}
Yicong Li$^{1,2,\star,\dagger}$, Yaron Meirovitch$^{3}$, Aaron T. Kuan$^{4}$, Jasper S. Phelps$^{4}$, Alexandra Pacureanu$^{6}$\\Wei-Chung Allen Lee$^{4,5}$, Nir Shavit$^{1}$, Lu Mi$^{1,\star}$
\end{tabular}\thanks{$^{\star}$ Corresponding authors: yicong\_li@g.harvard.edu, lumi@mit.edu}\thanks{$^{\dagger}$ Work done during the internship at MIT CSAIL.}\thanks{Supplementary material is available in this arXiv version.}}
\address{$^{1}$ Computer Science and Artificial Intelligence Laboratory, MIT\\
    $^{2}$ John A. Paulson School of Engineering and Applied Sciences, Harvard University\\
    $^{3}$ Department of Molecular and Cellular Biology, Harvard University\\
    $^{4}$ Department of Neurobiology, Harvard Medical School\\
    $^{5}$ F.M. Kirby Neurobiology Center, Boston Children’s Hospital, Harvard Medical School\\
    $^{6}$ ESRF, The European Synchrotron}
\begin{document}
%
\maketitle
\begin{abstract}
Comprehensive, synapse-resolution imaging of the brain will be crucial for understanding neuronal computations and function. In connectomics, this has been the sole purview of volume electron microscopy (EM), which entails an excruciatingly difficult process because it requires cutting tissue into many thin, fragile slices that then need to be imaged, aligned, and reconstructed. Unlike EM, hard X-ray imaging is compatible with thick tissues, eliminating the need for thin sectioning, and delivering fast acquisition, intrinsic alignment, and isotropic resolution. Unfortunately, current state-of-the-art X-ray microscopy provides much lower resolution, to the extent that segmenting membranes is very challenging. We propose an uncertainty-aware 3D reconstruction model that translates X-ray images to EM-like images with enhanced membrane segmentation quality, showing its potential for developing simpler, faster, and more accurate X-ray based connectomics pipelines.
\end{abstract}

\section{Introduction}
\label{sec:intro}

The field of connectomics investigates comprehensive maps of neuronal wiring and connectivity of the nervous system. Mapping the structure of neuronal networks is a key step toward understanding their function. Until recently, electron microscopy (EM) was the only approach that achieves comprehensive, nanometer-resolution imaging \cite{kasthuri2015saturated}. Unfortunately, obtaining three-dimensional (3D) EM volumes can be time and resource prohibitive. Because electrons are easily scattered, EM requires slicing massive numbers of thin sections and imaging them at nanoscale. An emerging imaging technique, X-ray holographic nano-tomography (XNH), has recently shown its potential to extract neuronal wiring, especially long-range myelinated axons \cite{kuan2020dense}. XNH can provide 3D imaging of whole specimens without nanoscale slicing during the imaging process, and thus significantly reduces the acquisition time and the potential existence of inter-slice alignment errors. Unfortunately, the speed and the scalability of XNH imaging on whole sample acquisition come, at least today, at the expense of delivering lower resolution images compared to EM. A large amount of time and resources could be saved if the reconstruction quality of XNH could be brought closer to that of EM.

\begin{figure*}[t]
\centering
\includegraphics[width=0.9\textwidth]{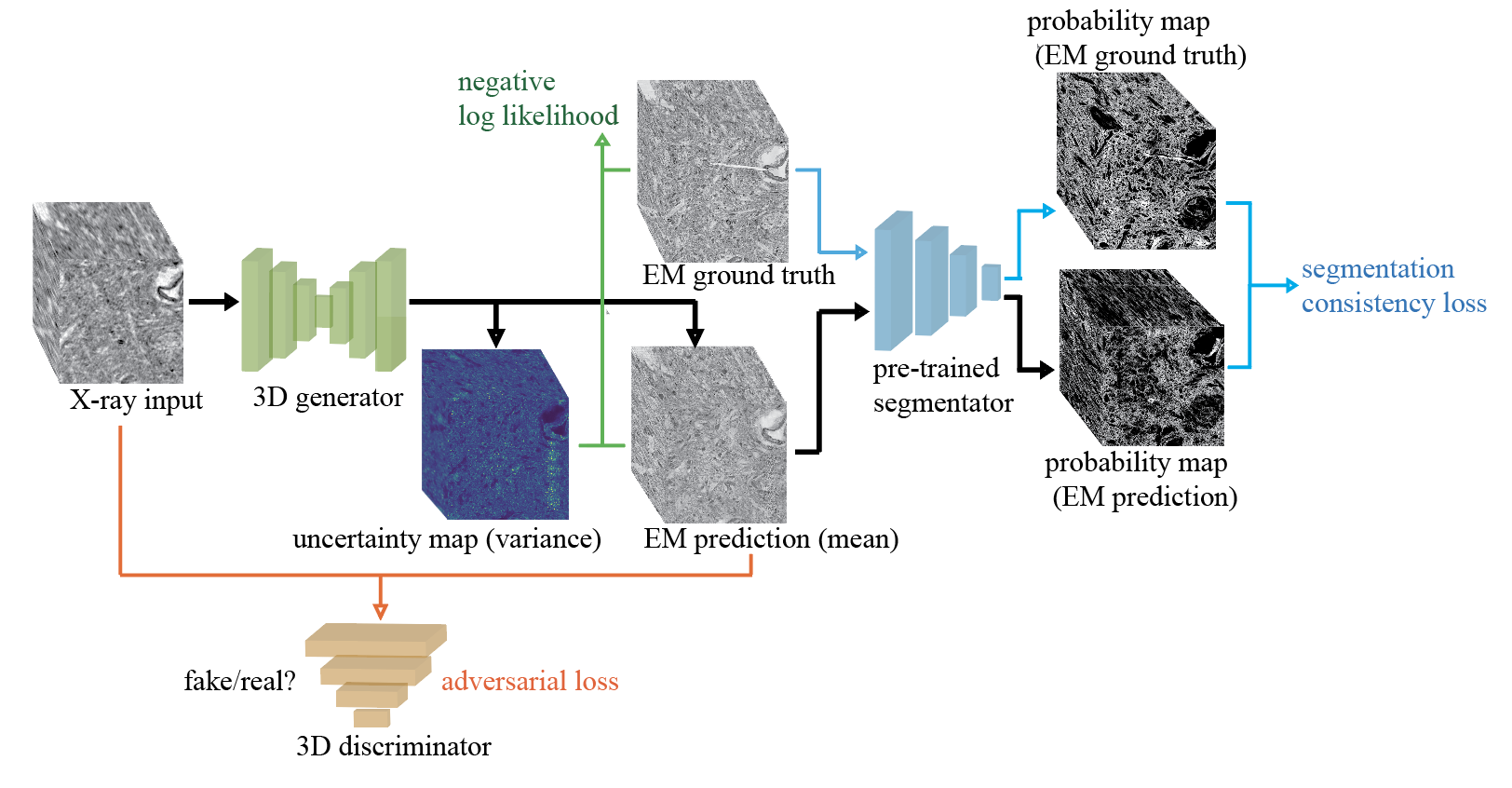}
\caption{Overview of the proposed framework. The 3D generator $G$ processes an image $x$ from the X-ray image domain $X$ to look similar to an image $y$ from the EM image domain $Y$ and outputs a corresponding uncertainty map. The 3D discriminator $D_Y$ learns to distinguish between a real EM image $y$ and a reconstructed output $G(x)$. The fixed pre-trained segmentation network $F_s$ measures the distance between the predicted membrane probability map $F_s(y)$ from the ground truth EM and the predicted membrane probability map $F_s(G(x))$ from the EM-like reconstruction. The model is optimized with a weighted combination of an adversarial loss, a negative log-likelihood (NLL), and a segmentation-consistency loss.}
\label{model}
\end{figure*}

Recent advances in deep learning based computer vision techniques \cite{lecun2015deep} have brought forth successful techniques that can assist in the above task. In the field of image-to-image translation, two pioneering works, \cite{isola2017image} and \cite{zhu2017unpaired}, develop 2D conditional generative adversarial networks (cGANs) \cite{mirza2014conditional} to translate images from one domain to the other. Such methods are further extended to 3D for processing medical data \cite{zeng2019hybrid,pan2018synthesizing,zhang2018translating}. Several other works have demonstrated the potential to use such techniques to further optimize the imaging process \cite{fang2021deep,wang2019deep,weigert2018content}.

However, to the best of our knowledge, such novel deep learning techniques have not previously been used to bridge the image quality gap between the EM and X-ray modalities in connectomics. We aim to use cross-modality image reconstruction to improve the quality of the X-ray image by leveraging its 3D consistency to achieve enhanced X-ray segmentation quality. Achieving this goal could substantially reduce the time needed for imaging large-scale volumes of densely stained neuronal circuits.

In this work, we make a pioneering first step in this direction by showing that a 3D cGAN can be used to reconstruct high-quality EM-like images from low-quality X-ray images, mitigating the trade-off between imaging time and resolution in connectomics. Our method allows us to reconstruct the details of neuronal borders (membranes) that are evident at the EM level but hardly visible in the original X-ray images. Major contributions of this work are as follows:
\begin{itemize}
    \item The first utilization of deep learning techniques to translate between the X-ray and EM modalities in connectomics.
    \item Our model demonstrates state-of-the-art reconstruction quality with 3D consistency and further improves the membrane segmentation performance.
    \item Our model also provides interpretable uncertainty maps during 3D reconstruction.
\end{itemize}
\section{Methods}
\label{method}

\subsection{Cross-modality image reconstruction framework}
As shown in Fig. \ref{model}, we utilize a cGAN combined with segmentation constraints to perform this image reconstruction task. Given a target EM image $y$ and an input X-ray image $x$, we build a 3D generator $G$, a 3D discriminator $D_Y$, and a fixed pre-trained segmentation network $F_s$. The 3D generator learns a mapping $X \rightarrow Y$ to reconstruct EM-like images from X-ray images as well as generates corresponding uncertainty maps and the 3D discriminator learns to distinguish the reconstructed EM-like images from the ground truth EM images. Afterward, the fixed pre-trained segmentation network will output segmentation probability maps of the ground truth and the reconstruction, the distance between which is minimized in order to provide additional structural constraints.

\begin{figure*}[t]
\centering
\includegraphics[width=0.85\textwidth]{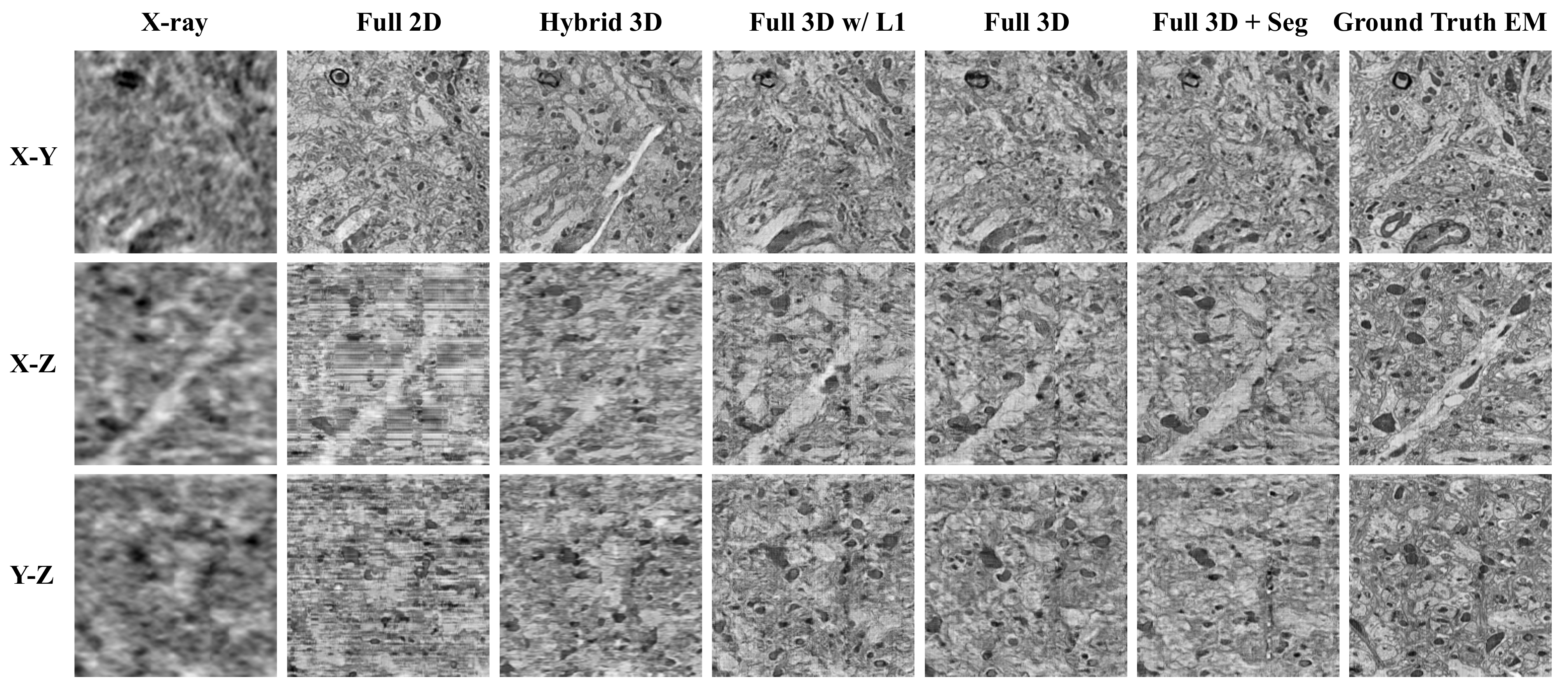}
\caption{Reconstruction results: visualization of images reconstructed by different methods along X-Y, X-Z Y-Z directions, in comparison with input X-ray and ground truth EM.}
\label{visual_result_1}
\end{figure*}

\subsection{Training objectives}

\subsubsection{Adversarial loss}
The adversarial loss is optimized as follows:
\begin{small}
\begin{equation}
\begin{split}
\label{adversarial loss}
       \mathop{min}_{G}\mathop{max}_{D_Y}\mathcal{L}_{\text{GAN}}(G, D_Y, X, Y) = \mathbb{E}_{y\sim p_{\text{data}}(y)}[\text{log}D_Y(y)] \\+ \mathbb{E}_{y\sim p_{\text{data}}(x)}[\text{log}(1-D_Y(G(x)))]
\end{split}
\end{equation}
\end{small}

\subsubsection{Negative log-likelihood}
The reconstruction model (generator) is optimized on aligned training pairs of X-ray and ground truth EM images $(x_i, y_i)$. Typically, we could have an additional $L_1$ loss term which further constrains the learning. Instead, to provide robustness analysis and interpretability via uncertainty estimation, we use the negative log-likelihood (NLL) \cite{kendall2017uncertainties} over an output Gaussian distribution $(G(x_i),\sigma_{y_i}^2)$ for every pixel $y_i$, where $G(x_i)$ indicates the reconstructed image, $\sigma_{y_i}^2$ indicates the measured uncertainty, $N^2$ is the number of pixels, and we have:
\begin{small}
\begin{equation}\label{NLL}
\mathcal{L}_{\text{NLL}}(G, X, Y) = \frac{1}{N^2}\sum_{i=1}^{N^2}\frac{\|y_i-G(x_i)\|_2^2}{2\sigma_{y_i}^2} + \frac{1}{2} \text{log}(2\pi \sigma_{y_i})
\end{equation}
\end{small}

\subsubsection{Segmentation-consistency loss} We utilize a pre-trained membrane segmentation network obtained from a related EM data domain \cite{kasthuri2015saturated}. It constrains the predicted probability maps from the ground truth EM image and the reconstructed EM-like image to be consistent in terms of membranes. The constraint is defined as:
\begin{small}
\begin{equation}\label{Seg}
\mathcal{L}_{\text{Seg}}(G, X, Y) = \|F_s(G(X))-F_s(Y)\|_2^2
\end{equation}
\end{small}

\subsubsection{Final objective}
Our final objective is a weighted combination of the adversarial loss $\mathcal{L}_{\text{GAN}}$, the negative log-likelihood $\mathcal{L}_{\text{NLL}}$, and the segmentation-consistency loss $\mathcal{L}_{\text{Seg}}$:
\begin{small}
\begin{equation}\label{seg_loss}
\begin{split}
\mathcal{L}(G, D_Y, X, Y) = w_{\text{GAN}}\mathcal{L}_{\text{GAN}}(G, D_Y, X, Y) \\+ w_{\text{NLL}}\mathcal{L}_{\text{NLL}}(G, X, Y) 
 + w_{\text{Seg}}\mathcal{L}_{\text{Seg}}(G, X, Y)
\end{split}
\end{equation}
\end{small}

\begin{figure*}[t]
\centering
\includegraphics[width=0.75\textwidth]{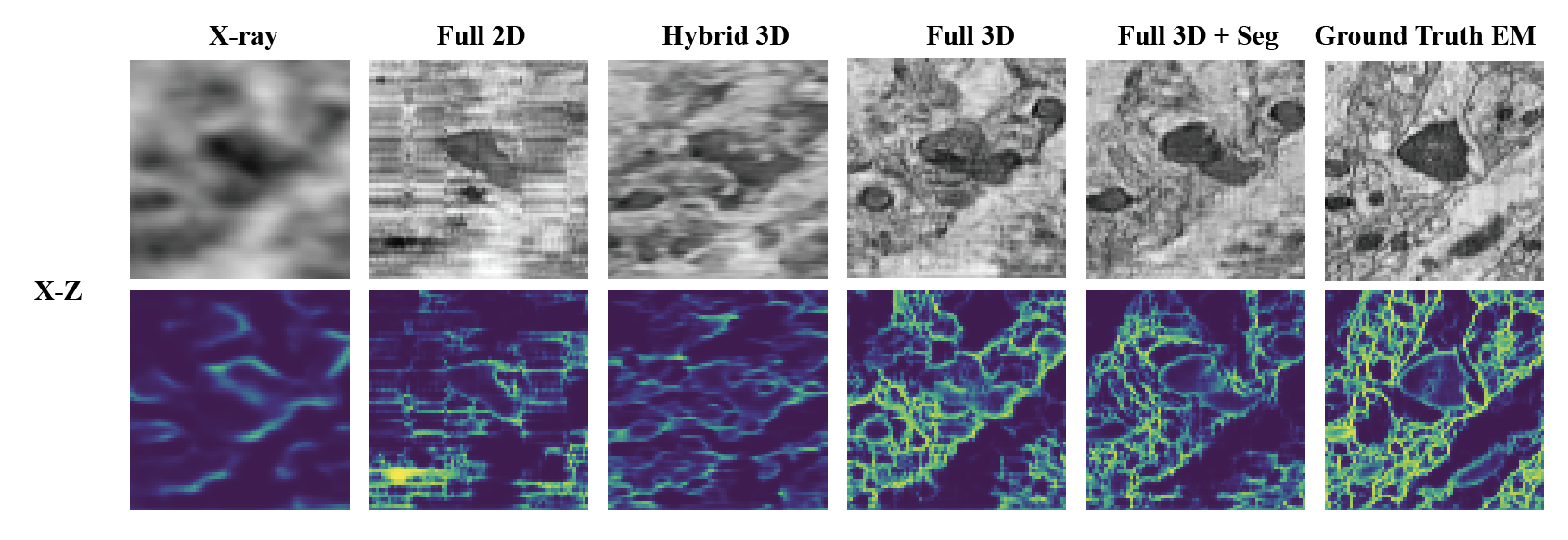}
\caption{Segmentation results: visualization of 2D segmentation probability map of membranes in the X-Z direction.}
\label{visual_result_2}
\vspace{-0.5cm}
\end{figure*}

\section{Experiments and Results}
\label{sec:pagestyle}

\begin{table}[t]
\setlength{\tabcolsep}{1.0pt}
\footnotesize
\centering
\begin{tabular}{c|c|cc|c|cc|c|cc}
\hline
{\color[HTML]{000000} \textbf{Model}} & {\color[HTML]{000000} \textbf{Dir.}} & {\color[HTML]{000000} \textbf{PSNR}}   & {\color[HTML]{000000} \textbf{SSIM}}  & \textbf{Dir.} & \textbf{PSNR}   & \textbf{SSIM}  & \textbf{Dir.} & \textbf{PSNR}   & \textbf{SSIM}  \\ \hline
{\color[HTML]{000000} Full 2D}        & {\color[HTML]{000000} }              & {\color[HTML]{000000} 14.670}          & {\color[HTML]{000000} 0.140}          &               & 14.490          & 0.124          &               & 14.500          & 0.106          \\
{\color[HTML]{000000} Hybrid 3D}      & {\color[HTML]{000000} }              & {\color[HTML]{000000} 14.701}          & {\color[HTML]{000000} 0.143}          &               & 14.566          & 0.129          &               & 14.702          & 0.108          \\
{\color[HTML]{000000} Full 3D w/ $L_1$}  & {\color[HTML]{000000} X-Y}           & {\color[HTML]{000000} 14.707}          & {\color[HTML]{000000} \textbf{0.153}} & X-Z           & 14.702          & {\ul 0.133}    & Y-Z           & 14.734          & {\ul 0.115}    \\
{\color[HTML]{000000} Full 3D}        & {\color[HTML]{000000} }              & {\color[HTML]{000000} \textbf{14.774}} & {\color[HTML]{000000} {\ul 0.150}}    &               & \textbf{14.763} & \textbf{0.134} &               & \textbf{14.793} & 0.114          \\
{\color[HTML]{000000} Full 3D + Seg}  & {\color[HTML]{000000} }              & {\color[HTML]{000000} {\ul 14.738}}    & {\color[HTML]{000000} 0.149}          &               & {\ul 14.722}    & 0.132          &               & {\ul 14.745}    & \textbf{0.116} \\ \hline
\end{tabular}
\caption{Quantitative evaluation on reconstruction quality. PSNR: peak signal-to-noise ratio; SSIM: structural similarity index measure; Dir.: direction; Bold Number: best score; Underlined Number: second best score.}
\label{table_result_1}
\end{table}

\begin{table}
\setlength{\tabcolsep}{1.0pt}
\centering
\scriptsize
\begin{tabular}{c|c|cc|c|cc|c|cc|c|cl}
\hline
{\color[HTML]{000000} \textbf{Model}} & {\color[HTML]{000000} \textbf{Dir.}} & {\color[HTML]{000000} \textbf{JS}}    & {\color[HTML]{000000} \textbf{Dice}}  & \textbf{Dir.}       & \textbf{JS}    & \textbf{Dice}  & \textbf{Dir.} & \textbf{JS}    & \textbf{Dice}  & \textbf{Dir.} & \textbf{JS} & \textbf{Dice} \\ \hline
{\color[HTML]{000000} X-ray}          & {\color[HTML]{000000} }                   & {\color[HTML]{000000} 0.366}          & {\color[HTML]{000000} 0.439}          & \multicolumn{1}{c|}{}    & 0.350          & 0.424          &                    & 0.361          & 0.429          &                    &0.167               &0.625  \\
{\color[HTML]{000000} Full 2D}        & {\color[HTML]{000000} }                   & {\color[HTML]{000000} \textbf{0.413}} & {\color[HTML]{000000} \textbf{0.546}} & \multicolumn{1}{c|}{}    & 0.382          & 0.495          &                    & 0.388          & 0.495          &                    &0.313         &0.838        \\
{\color[HTML]{000000} Hybrid 3D}      & {\color[HTML]{000000} X-Y}                & {\color[HTML]{000000} {\ul 0.403}}          & {\color[HTML]{000000} {\ul 0.521}}          & \multicolumn{1}{c|}{X-Z} & 0.369          & 0.466          & Y-Z                & 0.377          & 0.470          & 3D                 &0.279          & 0.824      \\
{\color[HTML]{000000} Full 3D}        & {\color[HTML]{000000} }                   & {\color[HTML]{000000} 0.399}          & {\color[HTML]{000000} 0.506}          & \multicolumn{1}{c|}{}    & {\ul 0.383}          & {\ul 0.498}          &                    & {\ul 0.392}          & {\ul 0.505}          &              &{\ul 0.424}  &{\ul 0.919}                   \\
{\color[HTML]{000000} Full 3D + Seg}  & {\color[HTML]{000000} }                   & {\color[HTML]{000000} 0.400}          & {\color[HTML]{000000} 0.507}          & \multicolumn{1}{c|}{}    & \textbf{0.386} & \textbf{0.502} &                    & \textbf{0.393} & \textbf{0.506} &                    &\textbf{0.433}   &\textbf{0.922}               \\ \hline
\end{tabular}
\caption{Quantitative evaluation on membrane segmentation. JS: Jaccard Score; Dice: Dice score; Dir.: direction; Bold Number: best score; Underlined Number: second best score.}
\label{table_result_2}
\end{table}

\subsection{Datasets}
\subsubsection{X-Ray2EM dataset}
For the EM modality, we have an EM dataset of mouse somatosensory cortex near the layer I/II boundary. The original voxel size of EM data is $4\times4\times45$ nm, while the volumes we used here have been downsampled by a factor of 16 in both X and Y directions to a voxel size of $64\times64\times45$ nm. The overall volume size is $2700\times2700\times243$ and is split into training/validation/test volumes with size of $2700\times1620\times243$, $2560\times512\times243$, and $2560\times512\times243$, respectively.

For the X-ray modality, we have an XNH dataset with a voxel size of $100\times100\times100$ nm. However, the true resolution is closer to $200\times200\times200$ nm \cite{kuan2020dense}. We warp this X-ray data into the EM volume space ($64\times64\times45$ nm) using an elastic alignment algorithm \cite{klein2009elastix,shamonin2014fast}. The overall volume size is also $2700\times2700\times243$. The training/validation/test splits are the same as the aforementioned EM dataset. It is important to note that there are larger resolution gaps and misalignment along the X-Z and the Y-Z directions than the X-Y direction between X-ray and EM modalities, leading to a more challenging reconstruction task along these two directions.

\subsubsection{Public EM segmentation benchmark dataset}
To pre-train the segmentation network, we use a related EM dataset of mouse cortex with corresponding membrane labels from Kasthuri. et al \cite{kasthuri2015saturated}.

\subsection{Performance evaluation}
We evaluate our proposed framework in comparison with two baseline methods on both the reconstruction task and the downstream segmentation task: (1) The original Pix2Pix model \cite{isola2017image}, which uses a 2D generator and a 2D discriminator. In this paper, we denote it as Full 2D and train with NLL instead of $L_1$ for a fair comparison. (2) A Hybrid 3D model consisting of a 3D generator and a 2D discriminator, similar to the method proposed in \cite{zeng2019hybrid}.

Fig. \ref{visual_result_1} shows the reconstruction results in three directions. Enhancement of spatial consistency can be observed from images reconstructed by Full 3D methods compared to the Full 2D and Hybrid 3D methods, particularly along the X-Z and Y-Z directions. Similarly in Fig. \ref{visual_result_2}, when the reconstructed EM images are evaluated by the downstream task, i.e., 2D membrane segmentation, the visual quality of the results generated by Full 3D methods is better than two baseline methods. Quantitatively, as shown in Table \ref{table_result_1} and \ref{table_result_2}, Full 3D methods outperform the Full 2D and the Hybrid 3D methods with considerable margins in most cases, in terms of PSNR and SSIM for reconstruction, as well as JS and Dice for segmentation. Meanwhile, Table \ref{table_result_1} shows that using NLL (Full 3D) instead of $L_1$ (Full 3D w/ $L_1$) not only provides uncertainty measurement but also enhances the reconstruction quality. And Table \ref{table_result_2} shows that using segmentation-consistency loss (Full 3D + Seg) can improve the segmentation quality.

\subsection{Uncertainty estimation}
\begin{figure}
\centering
\includegraphics[width=0.5\textwidth]{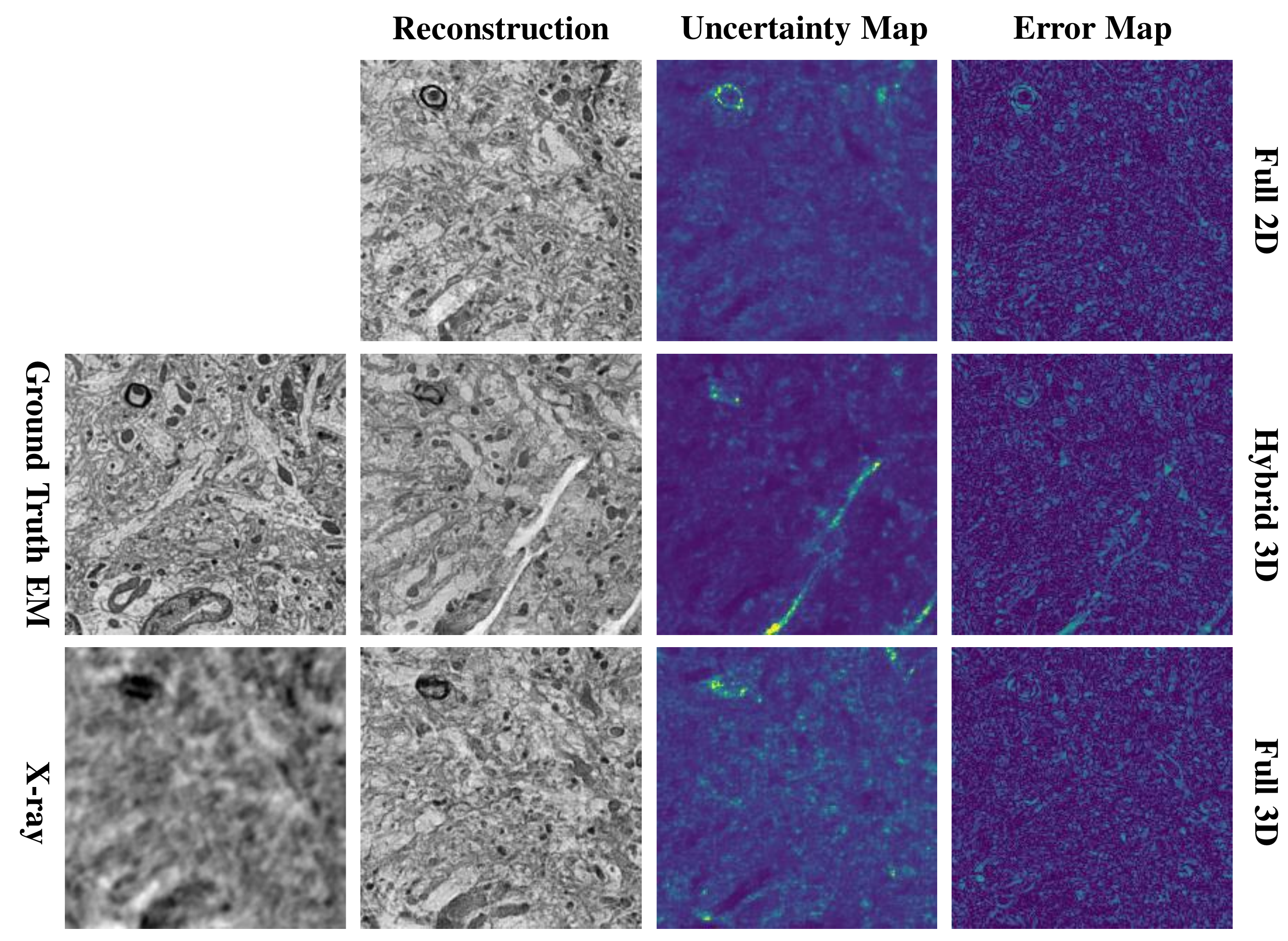}
\caption{Visualization of uncertainty estimation. Reconstruction \& Uncertainty: Gaussian output distribution (mean \& variance) generated by the models. Error: $L_1$ norm between the reconstruction and the ground truth EM.}
\label{visual_result_4}
\end{figure}

Intuitively, error maps are unable to capture visual ambiguity, and are not always accessible in real application scenarios. Uncertainty estimation, on the other hand, can highlight the uncertain areas that are consistent with the artifacts in the reconstructed images. For example, in the second row corresponding to Hybrid 3D in Fig. \ref{visual_result_4}, the model clearly hallucinates a long and slender object in the reconstruction which is not shown in the ground truth EM. The uncertainty map correctly highlights this artifact, providing a richer interpretation than the error map.

\section{Conclusions}
\label{sec:typestyle}

In this work, we propose a cross-modality reconstruction approach to reconstruct high-quality EM-like images from low-quality X-ray images for connectomics. We find that the Full 3D model is the key to improving the reconstruction and membrane segmentation results. Meanwhile, we also show the utilization of negative log-likelihood could improve the model performance as well as provide interpretable uncertainty maps. We believe our proposed approach will potentially allow the reconstruction of large-scale brain structures achieved in a fast and scalable way.

\section{Compliance with ethical standards}
\label{sec:ethics}
This research study was conducted retrospectively using animal subject data made available in open access by \cite{kuan2020dense}. Ethical approval was not required as confirmed by the license attached with the open access data.

\vspace{-0.05cm}
\section{Acknowledgments}
\label{sec:acknowledgments}
This research is funded by the NSF (IIS-1607189, CCF-1563880, IOS-1452593, 1806818) and the NIH (EB032217, MH117808, MH128949). Aaron T. Kuan is supported by NIH (EB032217), Wei-Chung Allen Lee is supported by NIH (MH128949), Lu Mi is supported by a Mathworks Fellowship, and Yaron Meirovitch is supported by NIH (5U24NS109102) \& NIH (U01 NS108637). This work is also supported by the European Research Council (ERC) under the European Union’s Horizon 2020 research and innovation programme (grant agreement n°852455). We acknowledge the ESRF for granting beamtime for proposal LS2892.

\bibliographystyle{IEEEbib}
\bibliography{refs}

\section{Supplementary Material}

\subsection{Visualization of domain differences}
\begin{figure}[htb]
\centering
\includegraphics[width=0.45\textwidth]{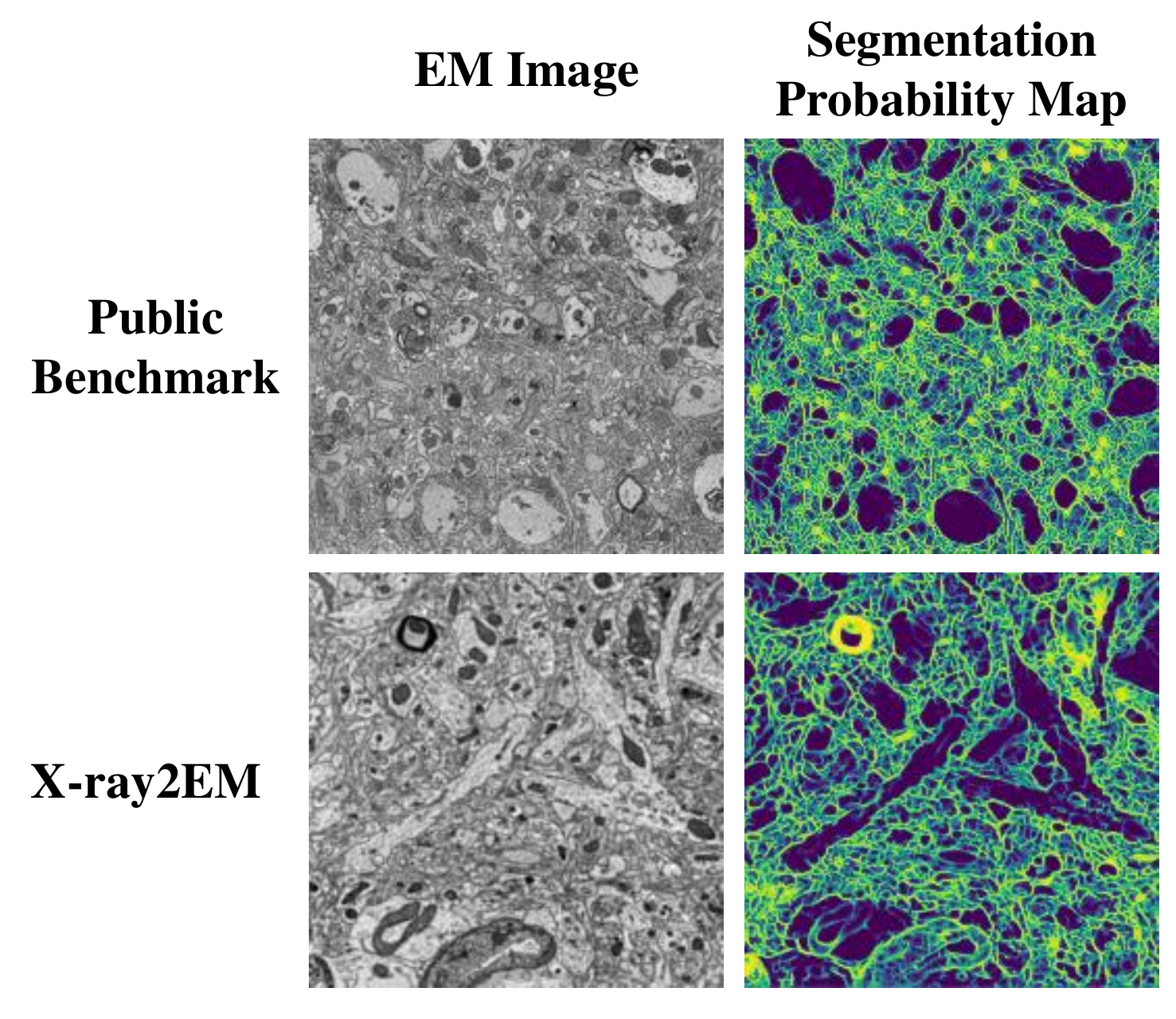}
\caption{Comparison of domain differences between benchmark EM dataset and X-Ray2EM dataset.}
\label{supp_visual_result_2}
\end{figure}

In this paper, we use a segmentation network pre-trained on a public benchmark EM membrane segmentation dataset to provide segmentation consistency constraints, in the absence of segmentation labels in X-Ray2EM dataset. Both datasets are EM images of mouse cortex and we show in Fig. \ref{supp_visual_result_2} that the domain difference between these two datasets has a negligible negative effect on the segmentation results.

\subsection{Additional 2D membrane segmentation results}
\begin{figure*}[htb]
\centering
\includegraphics[width=0.85\textwidth]{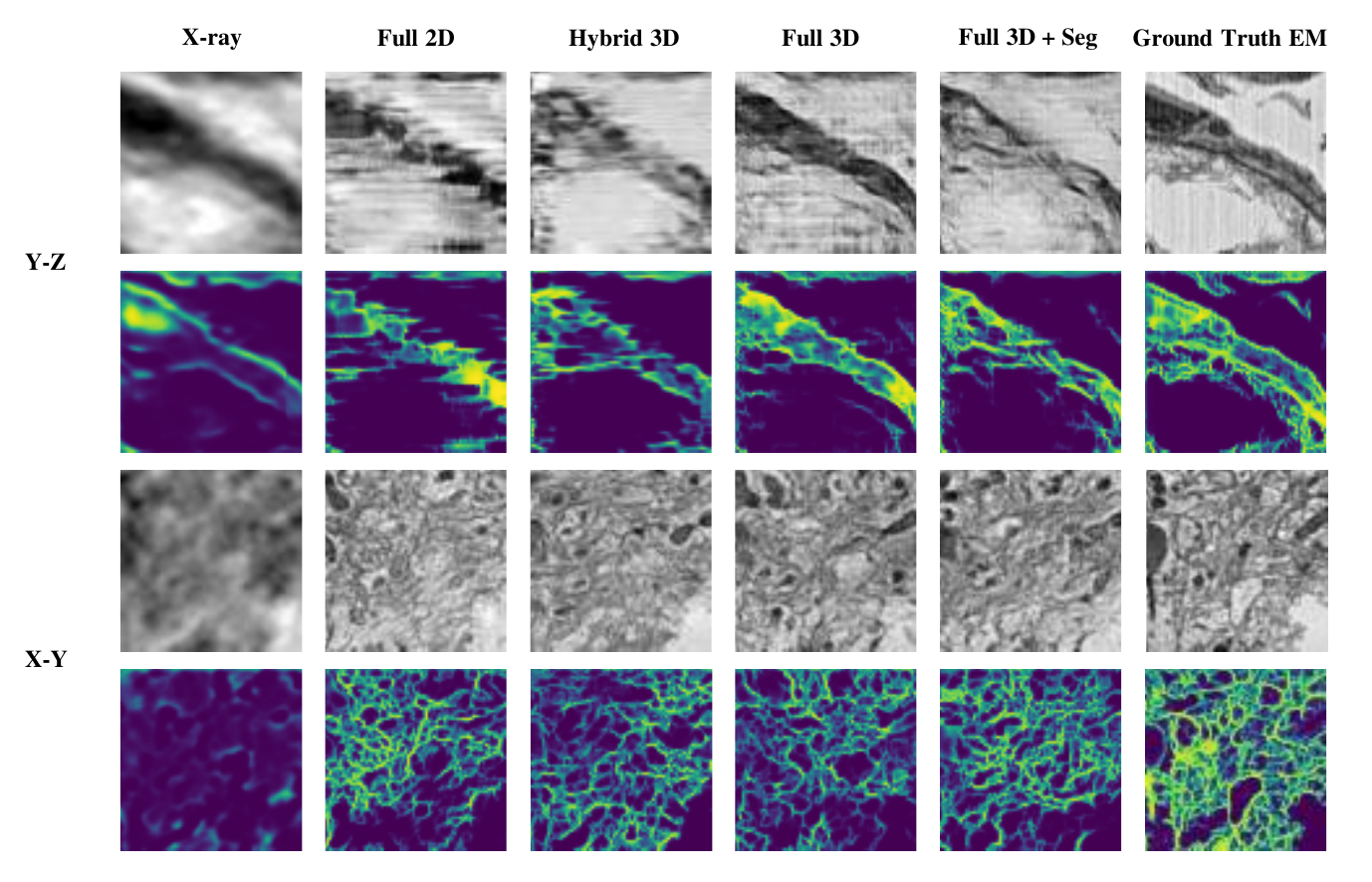}
\caption{Segmentation results: visualization of 2D segmentation probability map of membrane along the Y-Z and the X-Y directions.}
\label{supp_visual_result_1}
\end{figure*}

In Fig. \ref{supp_visual_result_1}, the reconstructed EM images are evaluated by downstream tasks, i.e., 2D membrane segmentation. It is consistent with the results (X-Z direction) in the main paper that along Y-Z and X-Y directions, the segmentation quality of EM reconstructed by Full 3D method is also better than the two baseline methods.

\subsection{Ablation studies}
\subsubsection{$L_1$ loss v.s. NLL}
In our work, we use NLL as one of the training objectives instead of $L_1$ loss. Results in the main paper show that the reconstruction results of Full 3D model trained with NLL are better than or comparable with that of Full 3D w/ $L_1$ model in most cases, indicating that using NLL for training not only provides the estimation of uncertainty but can also improve the reconstruction performance.

\subsubsection{W/ seg v.s. w/o seg}
According to the results shown in the main paper, Full 3D + Seg method achieve comparable results to Full 3D w/o Seg on the reconstruction task and prevails on the 2D membrane segmentation task, indicating that features like neuronal boundaries are better preserved by applying segmentation-consistency constraints.

\subsection{Software and hardware requirements}
All experiments in the paper are implemented in Python 3.6.12 using PyTorch 1.8.0 framework on a computation server with a 2.30GHz Intel(R) Xeon(R) Gold 5218 CPU and a GeForce RTX 3090 GPU.

\subsection{Segmentation network pre-training}
The segmentation network has a U-Net architecture originated from \cite{ronneberger2015u}. We use a cross-entropy loss function and train the network for 3000 iterations with a batch size of 4. Adam \cite{kingma2014adam} optimizer is used with an initial learning rate of 0.0001 and a weight decay of 0.00005.

\subsection{Reconstruction framework training}
The generator has a U-Net like architecture and the discriminator is a PatchGAN, both are 3D versions of the 2D Pix2Pix model proposed in \cite{isola2017image}. Input to the generator is randomly cropped from the whole volume to a size of $128\times128\times64$ and the output reconstruction also has the same size. As for the pre-trained 2D segmentation network following the generator, we randomly select one section from each of the three directions of the reconstructed EM volume to perform the inference process, leading to three segmentation probability maps. Only the generator and the discriminator participate in the training process, and the parameters of the segmentation network are fixed. The whole framework is also trained using Adam optimizer in an end-to-end fashion for 100 epochs, with a constant learning rate of 0.0002 for the initial 50 epochs and linearly decayed to 0 for the remaining 50 epochs. The batch size is 1. Loss function weights $w_{\text{NLL}}$, $w_{\text{GAN}}$, and $w_{\text{Seg}}$ are 0.00002, 1.0, and 1.0, respectively.

\subsection{X-Ray2EM dataset}
For the EM modality, we have a serial section TEM dataset. For the X-ray modality, we have an X-ray holographic nanotomography dataset \cite{kuan2020dense}. These datasets can be found at:
\begin{itemize}
    \item \textcolor{purple}{https://www.lee.hms.harvard.edu/kuan-phelps-et-al-2020}
\end{itemize}

\subsection{Public EM segmentation benchmark dataset}
To pre-train the segmentation network, we use a serial section SEM dataset of mouse cortex with corresponding membrane labels from Kasthuri. et al \cite{kasthuri2015saturated}. This dataset can be found at:
\begin{itemize}
    \item \textcolor{purple}{https://lichtman.rc.fas.harvard.edu/vast/}
\end{itemize}

\end{document}